\documentclass[12pt]{article}
\usepackage{amsfonts}
\usepackage{bbm}
\usepackage{bm}
\usepackage{amsmath}
\usepackage{epsfig}
\usepackage{amssymb,lscape}
\usepackage[matrix,arrow,curve]{xy}
\usepackage[english]{babel}

\allowdisplaybreaks

\numberwithin{equation}{section}

\newcommand{\bmat}{\left(\begin{array}}
\newcommand{\emat}{\end{array}\right)}

\def\gtrsim{\mathrel{\raise.3ex\hbox{$>$\kern-.75em\lower1ex\hbox{$\sim$}}}}

\def\be{\begin{equation}}
\def\ee{\end{equation}}
\def\bea{\begin{eqnarray}}
\def\eea{\end{eqnarray}}

\newcommand{\ha}{\hat\alpha}
\newcommand{\hb}{\hat\beta}
\newcommand{\hc}{\hat\gamma}
\newcommand{\hd}{\hat\delta}
\newcommand{\hr}{\hat\rho}
\newcommand{\hs}{\hat\sigma}
\newcommand{\Vp}{V^{\prime}}
\newcommand{\Vpp}{V^{\prime\prime}}
\newcommand{\vp}{v^{\prime}}
\newcommand{\vpp}{v^{\prime\prime}}
\newcommand{\wpr}{\omega^{\prime}}
\newcommand{\wpp}{\omega^{\prime\prime}}
\newcommand{\sip}{\sigma^{\prime}}
\newcommand{\sipp}{\sigma^{\prime\prime}}
\newcommand{\tp}{\tau^{\prime}}
\newcommand{\tpp}{\tau^{\prime\prime}}

\begin{document}
\pagestyle{plain}
\begin{titlepage}
\begin{center}
\large{\bf On the exceptional generalised Lie derivative for $d\geq7$ \\[4mm]}
\large{J. A. Rosabal${}^{a,b}$
 \\[4mm]}
\small{
${}^a${\em Departamento de F\'isica, Universidad de Buenos Aires CONICET-UBA.} \\[-0.3em]
{\em Pabell\'on I, Ciudad Universitaria, Buenos Aires, Argentina.}\\
[0.3cm]
${}^b${\em Institut de Physique Th\'eorique,
CEA/ Saclay \\
91191 Gif-sur-Yvette Cedex, France}  \\
[0.3cm]
{\verb"arosabal@df.uba.ar"}\\
[0.5cm]}
\small{\bf Abstract} \\[0.3cm]
\end{center}

In this work we revisit the $E_8\times\mathbb{R}^{+}$ generalised Lie derivative encoding the algebra of diffeomorphisms and gauge transformations of compactifications of M-theory on eight-dimensional manifolds, by extending certain features of the $E_7\times\mathbb{R}^{+}$ one. Compared to its $E_d\times\mathbb{R}^{+},\ d\le 7$ counterparts, a new term is needed for consistency. However, we find that no compensating parameters need to be introduced, but rather that the new term can be written in terms of the ordinary generalised gauge parameters by means of a connection. This implies that no further degrees of freedom, beyond those of the field content of the $E_{8}$ group, are needed to have a well defined theory. We discuss the implications of the structure of the $E_8\times\mathbb{R}^{+}$ generalised transformation on  the construction of the $d=8$ generalised geometry. Finally, we suggest  how to lift the generalised Lie derivative to eleven dimensions.
\vfill


\end{titlepage}

\newpage
\setcounter{page}{1}
\pagestyle{plain}
\renewcommand{\thefootnote}{\arabic{footnote}}
\setcounter{footnote}{0}

\tableofcontents

\newpage

\section{Introduction}

One of the most remarkable property of string theory is, perhaps, that its non linear sigma model formulation on different backgrounds
may define the same string theory at the quantum level. This property is known as ``duality''. The transformations between equivalent backgrounds
can be packaged in some groups of gauge symmetry. It is well known, for instance, that $D=11$ supergravity, or the effective action of type II string theory,   compactified on a $d$ torus $T^d$ ($T^{(d-1)}$) has an $E_{d}(\mathbb{R})$ duality, for $d\leq10$, see
\cite{Hull:1994ys} and references therein. For the full compactified theory on $T^{11}$ it is conjectured that $E_{11}$ could be the duality group \cite{West:2001as}, which actually could be a symmetry of the full M-theory, independently of the compactification \cite{Riccioni:2007au}.

The bosonic $D=10$ supergravity, whose field content is the metric $g_{\mu\nu}$ the Kalb-Ramond field $B_{\mu\nu}$ and the dilaton $\varphi$, has a non manifest $O(10,10)$ symmetry. This symmetry is known as T-duality. The T-dual covariant description of this supergravity theory is based on two different but related approaches. On the one hand double field theory (DFT) \cite{Hull:2009mi}, earlier versions of DFT can be found in \cite{Duff:1989tf}, \cite{Tseytlin:1990nb}, \cite{Tseytlin:1990va}, \cite{Siegel:1993xq} and \cite{Siegel:1993th}, which describes string backgrounds in terms of fields on a doubled twenty dimensional  space transforming under the $O(10,10)$ group. On the other hand generalised geometry \cite{Hitchin:2004ut}, \cite{Coimbra:2011nw} which unifies the local diffeomorphisms and gauge transformations of the $2$-form on a generalised tangent space $TM\oplus T^{*}M$ which has a natural $O(10,10)$ structure.

In DFT all fields and parameters are required to satisfy the section condition or strong constraint. This implies that they depend on only ten coordinates and, therefore, locally DFT is equivalent to generalised geometry. The strong constraint can be locally relaxed \cite{Grana:2012rr}, \cite{Dibitetto:2012rk},  \cite{Aldazabal:2013sca}. Some works in the direction of understanding, from the world sheet perspective, the origin of the strong constraint and its possible relaxation have appeared in the literature \cite{Blumenhagen:2014gva}, \cite{Betz:2014aia}, \cite{Cederwall:2014kxa}. However, the geometric interpretation of this relaxed theory is still not clear.

In the context of the exceptional groups the generalised geometry approach was first presented in \cite{Hull:2007zu} to describe the U-duality invariant $E_{d}$ theories with $d\leq7$ and later developed in a series of papers \cite{Pacheco:2008ps},  \cite{Coimbra:2011ky} and \cite{Coimbra:2012af}.
The extended field theory, as the counterpart of double field theory, was first presented in \cite{Berman:2010is}, \cite{Berman:2011pe}, for $d=4,5$ and for $d=6,7$ using the $E_{11}$ non linear formalism in  \cite{Berman:2011jh}. The geometric counterpart of DFT for the $E_7$ group in $d=7$ was developed in \cite{Aldazabal:2013mya}. In this work the relation between the four dimensional gauged maximal supergravity theory \cite{de Wit:2007mt}  and the U-duality extended $E_{d}\times{\mathbb{R}^{+}}$ theory was pointed out. More recent applications of the  $E_{d}\times{\mathbb{R}^{+}}$ generalised geometry  can be found in \cite{Coimbra:2014qaa}, \cite{Lee:2014mla}. For other extensions and applications of DFT and the extended field theory see \cite{Baron:2014yua},  \cite{Berkeley:2014nza}, \cite{Bedoya:2014pma} and \cite{Berman:2014jsa}.

For $d\geq8$, some extended-like works have appeared in the literature. In  \cite{Godazgar:2013rja}, using the non linear realization of the $E_8$ group, the authors were able to write the supergravity action restricted to eight dimensions including the dual graviton field but not the gauge transformations of this. In \cite{Aldazabal:2013via}, starting from the extended $E_d\times\mathbb{R}^{+}$ generalised Lie derivative, with $d\leq7$, it was attempted to complete it
using the tensor hierarchy mechanism \cite{deWit:2005hv},  \cite{deWit:2005ub}, \cite{deWit:2008ta}, to get the eleven dimensional transformation. Also, it was shown that even the generalised transformations, beyond seven dimensions, have a gauge structure, \footnote{See \cite{Berman:2012vc} for a discussion about it for all $E_{d}$ with $d\leq7$.} some obstructions of consistency and covariance came out at some given level (at the level of the tensor hierarchy which corresponds to the adjoint representation of the $E_d$ group). This is the reason why for the $E_8$ group the algebra of the generalised Lie derivative does not close when naively
the $E_d$-series ($d\leq7$) of generalised Lie derivatives is extended to $E_8$ \cite{Berman:2012vc}. For the $E_8$ group the fundamental and the adjoint representations are essentially the same representation.

In this line, what is called Exceptional Field Theory has been developed in \cite{Hohm:2013vpa}, \cite{Hohm:2013uia} and \cite{Hohm:2014fxa}. This theory uses the $E_d\times\mathbb{R}^{+}$ with $d\leq8$ gauge transformation but embedded in eleven dimensions. The tensor hierarchy mechanism is not enough to achieve the closure of the algebra, thus a mysterious compensating field has to be added to compensate its failure. This new parameter does not fit in the fundamental representation of the algebra of the exceptional groups. Hence, this parameter lies in a new direction on the extended space, which is equivalent to say that the $E_d$-generalised tangent space gets larger. Some issues of this approach are discussed in section \ref{e8rmas}.

In $d=8$, the dual gravity and higher dual fields become relevant. The dual graviton, for instance, is described through a field with a mixed symmetry $A_{(1,8)}$ whose gauge parameter is a mixed symmetry tensor $\tau_{(1,7)}$. The conventional gauge field theories seem not to work for this kind of fields. For this reason a consistent generalised geometry or extended description based on the $E_{8}\times{\mathbb{R}^{+}}$ group can not be found yet in the literature.

In this paper we present the $E_{8}\times{\mathbb{R}^{+}}$ generalised Lie derivative. This one could be the base for establishing the generalised geometry description of the $d=8$ U-duality theory and perhaps going beyond the $d=8$ case. The generalised Lie derivative, in components, is given by
\be
(\hat{\cal L}_{\xi}V)^M=\xi^P\partial_PV^M-f_A{}^{M}{}_Nf^{AP}{}_Q\partial_P\xi^QV^N+\partial_P\xi^PV^M-f^{MP}{}_Q\Sigma_PV^Q.
\ee
The crucial difference with \cite{Hohm:2014fxa} is that $\Sigma$ is not an independent parameter but is given by
\be
\Sigma_{P}=\frac{1}{60}f_J{}^K{}_L\tilde{\Omega}_{P K}{}^L\xi^J,
\ee
where $f_J{}^K{}_L$ are the structure constants of $\mathfrak{e}_8$  and the derivative only has components in eight directions, as well as the field $\tilde{\Omega}_{P K}{}^L$ in the index $P$. Actually,  this field is a generalised connection on the $8$-dimensional manifold, see \cite{Coimbra:2011ky} for the definition of the generalised connection in this context.

On the other hand, all gauged maximal supergravity theories in lower dimensions can be constructed making no references to string theory or $D=10,\ 11$ supergravity.
This kind of theories are consistent upon the tensor hierarchy mechanism. The most general theories of gauged maximal supergravity are those where the trombone symmetry is gauged \cite{LeDiffon:2008sh}. In these theories the gaugings are distributed over some given representations of the exceptional groups, including the fundamental one. Comparing to those supergravity theories that come from  a reduction of the $D=11$ supergravity, it seems, the former have more gaugings than the ones that can be obtained from the latter.

This mismatch can be fixed adding more fields than those that $D=11$ supergravity has. In fact, adding to it an infinite number of fields, starting at the fourth level of the $E_{11}$ algebra, it is possible to get all gauged maximal supergravity theories \cite{Riccioni:2009xr}, \cite{Riccioni:2010xx}.

In this paper we will focus on the three dimensional gauged maximal supergravity whose duality group is $E_{8}\times\mathbb{R}^{+}$, where the $\mathbb{R}^{+}$ factor is associated with the trombone symmetry. We start by establishing  the correspondence between the gaugings (fluxes) and the generalised Lie derivative in eight dimensions. Concretely, the fluxes are defined as the coefficients of the expansion of the generalised vector $\hat{\cal L}_{E_{\bar{A}}}E_{\bar{B}}$ in the frame $E_{\bar{C}}$, namely
\be
\hat{\cal L}_{E_{\bar{A}}}E_{\bar{B}}=F_{\bar{A}\bar{B}}{}^{\bar{C}}E_{\bar{C}}\label{flu}
\ee
which holds for all $E_{d}\times\mathbb{R}^{+}$ with $d\leq7$. We will proceed by assuming ($\ref{flu}$) also holds for $d=8$. Finally, we will suggest how the $d=8$ generalised transformation can be lifted to eleven dimensions.

The paper is organised as follows. In section \ref{e7rmas} we  make a summary of some previous results regarding the $E_7\times \mathbb{R}^{+}$ generalised Lie derivative and the fluxes. Then, with the aim of making contact with the generalised geometry approach, we present the $SL(8)$ and $SL(7)$ decomposition of the $E_7$ group and the generalised Lie derivative. In section \ref{e8rmas} we review the known approaches regarding the $E_8\times \mathbb{R}^{+}$ generalised transformation. We introduce a detailed $E_8$ group-theoretic analysis and then perform the $SL(9)$ and $SL(8)$ decomposition of the known $d=8$ generalised transformation. Based on the decomposition and the lessons learnt from the $E_7\times \mathbb{R}^{+}$  case we move to the construction of the $E_8\times \mathbb{R}^{+}$ generalised Lie derivative and then we check its consistency and compatibility. The section \ref{e11} is dedicated to discuss the possible lifting of the $d=8$ generalised transformation to eleven dimension. The summary and conclusions are presented in section \ref{conclu}.

\section{$E_7\times \mathbb{R}^{+}$}\label{e7rmas}

\subsection{Summary of previous results}

 In this section we are interested in exploring the extended $E_7\times \mathbb{R}^{+}$ generalised transformation and the obstructions to lift it to eleven dimensions. The obstructions, for doing this extension, come from some ambiguities in the writing of the dual diffeomorphism in $d=7$ and the closure of the algebra. We will show that they can be avoided if instead of starting with the extension from $E_7\times \mathbb{R}^{+}$ one starts from the $E_8\times \mathbb{R}^{+}$ group in $d=8$. Discussions on how to build the $E_8\times \mathbb{R}^{+}$ extended transformations and the implications for the generalised geometry are presented in the next section.

Our starting point will be the generalised $E_7\times \mathbb{R}^{+}$  transformation  \cite{Berman:2012vc}, \cite{Aldazabal:2013mya} which in a given local generalised patch reads
\be
(\delta_{\xi}V)^M=({\cal L}_{\xi}V)^M=\xi^P\partial_PV^M-A^M{}_N{}^P{}_Q\partial_P\xi^QV^N=(L_{\xi}V)^M+Y^M{}_N{}^P{}_Q\partial_P\xi^QV^N\label{transformation}
\ee
where $V$ and $\xi$ are generalised vectors. All $E_7\times \mathbb{R}^{+}$ generalised vectors are weighted such that $V=e^{-\Delta}\tilde{V}$, being $\tilde{V}$ a pure $E_7$ generalised vector and $e^{-2\Delta}=det(e)$. $M=1\ldots 56$ is an index in the representation space of $E_7$ (fundamental representation). $L_{\xi}$ is the ordinary Lie derivative on the generalised $E$-tangent bundle, that locally can be seen as
\be
E\simeq TM\oplus \Lambda^2T^*M\oplus \Lambda^5T^*M\oplus (\Lambda T^*M\otimes\Lambda^7T^*M)\label{tangent},
\ee
\be
A^M{}_N{}^P{}_Q=12P_{(adj)}{}^M{}_N{}^P{}_Q-\frac{1}{2}\delta_N^M\delta_Q^P,
\ee
\be
Y^M{}_N{}^P{}_Q=-A^M{}_N{}^P{}_Q+\delta_Q^M\delta_N^P\label{Y}
\ee
and
\be
P_{(adj)}{}^M{}_N{}^P{}_Q=K_{ab}(t^a)_N{}^M(t^b)_Q{}^P\qquad a=1\ldots 133 \label{pro-adj}
\ee
is a projector $P_{(adj)}:{\bf56}\times{\bf56}^{*}\rightarrow{\bf 133}=adj(E_7)$
being $t^a$ the generators of the algebra $\mathfrak{e}_7$  and $K_{ab}$ the Cartan-Killing metric. In terms of the seven dimensional objects the generalised vector can be identified with
\be
V=(v\ , \  \omega_2 \ , \ \sigma_5 \ , \ \tau_{(1,7)})\label{vector}
\ee
where $v$ is a vector, $\omega_2$ is a 2-form, $\sigma_5$ is a 5-form and $\tau_{(1,7)}$ is a tensor with a mixed symmetry.
The transformation (\ref{transformation}) satisfies the relation
\be
\big[{\cal L}_{\xi_1},{\cal L}_{\xi_2}\big]V={\cal L}_{{\cal L}_{\xi_1}\xi_2}V\label{algebra},
\ee
provided the section condition holds
\be
\Omega^{PQ}\partial_P\otimes\partial_Q=P_{(adj)}{}_{MN}{}^{PQ}\partial_P\otimes\partial_Q=0\label{sc}
\ee
which implies
\be
Y^M{}_N{}^P{}_Q\partial_M\otimes\partial_P=0,
\ee
where $\Omega^{PQ}$ is the symplectic invariant. The relation (\ref{algebra}) ensures the Leibniz property and thus the covariance of the generalised Lie derivative. The most effective way to see that, is writing the generalised Lie derivative as the bracket
\be
{\cal L}_{\xi}=\big[\xi,\ \big].\label{bracket}
\ee
Then it is easy to see that the Leibniz property implies
\be
\big[\xi_1,\big[\xi_2,V\big]\big]=\big[\big[\xi_1,\xi_2\big],V\big]+\big[\xi_2,\big[\xi_1,V\big]\big]\label{leibnitz}
\ee
which actually is
\be
\big[\xi_1,\big[\xi_2,V\big]\big]-\big[\xi_2,\big[\xi_1,V\big]\big]=\big[\big[\xi_1,\xi_2\big],V\big]\equiv \big[{\cal L}_{\xi_1},{\cal L}_{\xi_2}\big]V={\cal L}_{{\cal L}_{\xi_1}\xi_2}V. \label{leibnitz1}
\ee

There are at least two known solutions of the section condition \cite{Aldazabal:2013mya}. Here, we are interested in making contact with the generalised geometry approach \cite{Coimbra:2011ky},  for this reason we only focus on the $SL(8)$ decomposition of (\ref{transformation}). In this decomposition,  the derivative $\partial_{\alpha}$, can be viewed as an object (a section) of $E^{*}$ (the dual generalised tangent bundle) through the embedding
\be
\partial_M=\big(\partial_{\alpha}=\frac{1}{2}\partial_{[\alpha 8]}\quad,\quad \partial_{[\alpha,\beta]}=0 \quad,\quad \partial^{[\hat{\alpha},\hat{\beta}]}=0\big)
\ee
where the hatted indices run from $1$ to $8$ while the unhatted ones run from $1$ to $7$. This embedding is a solution of (\ref{sc}).

Given a generalised parallelisable manifold \cite{Lee:2014mla} it is possible to pick a global generalised frame $E_{\bar{A}}$ on the generalised tangent space and define the so called generalised fluxes \footnote{In general $F_{\bar{A}\bar{B}}{}^{\bar{C}}$ is not a constant.},
\be
{\cal L}_{E_{\bar{A}}}E_{\bar{B}}=F_{\bar{A}\bar{B}}{}^{\bar{C}}E_{\bar{C}}.\label{fluxes}
\ee
Using (\ref{transformation}) it is possible to prove that $F_{\bar{A}\bar{B}}{}^{\bar{C}}$ belongs to the representations dictated by gauged supergravity \cite{de Wit:2007mt}, $F\in\bf{56+912}$.

In what follows we display the proof that $F_{\bar{A}\bar{B}}{}^{\bar{C}}$ is in the $\bf{56+912}$ representations of the $E_7$ group. The interest in doing this, is because of one may take advantage of this result to propose a general principle to build the generalised transformations which holds, not only for $E_d\times\mathbb{R}^{+}$ with $d\leqslant7$, but also for the $E_8\times\mathbb{R}^{+}$ group and perhaps for $d\leqslant11$.

In curved indices, the fluxes can be written as \cite{Aldazabal:2013mya}
\be
F_{MN}{}^P=\Omega_{MN}{}^P-12P_{(adj)}{}^P{}_N{}^R{}_S\Omega_{RM}{}^S+\frac12\Omega_{RM}{}^R\delta_N^P\label{fluxescurv}
\ee
where
\be\label{weitzenbock}
\Omega_{MN}{}^P=E_N{}^{\bar{A}}\partial_ME_{\bar{A}}{}^P=\Omega_{M}{}^0(t_0)_N{}^P+\tilde{\Omega}_{M}{}^a(t_a)_N{}^P
\ee
with $\Omega_{M}{}^0=\partial_M\Delta$ and $(t_0)_N{}^P=-\delta_N{}^P$ is the generator in $\mathbb{R}^+$. In terms of the projectors $P_{(R_i)}:{R_1\times R_2}\rightarrow R_i$, where $\bf{R_1=56}$, $\bf{R_2=133}$  and $\bf{R_3=912}$ are the three first irreducible representations of the $E_7$ group \cite{deWit:2002vt},
\bea\label{projectors}
P_{(56)M}{}^{a \ R}{}_b & = &\frac{8}{19}\,(t^a)_M{}^{K}\,(t_b)_K{}^{R}\\ \nonumber
P_{(912)M}{}^{a \ R}{}_{b} & = &\frac{1}{7}\delta_M^{R}\delta^a_b -\frac{12}{7}\,(t^a)_K{}^{R}\,(t_b)_M{}^{K} + \frac4{7}(t^a)_M{}^{K}\,(t_b)_K{}^{R},
\eea
(\ref{fluxescurv}) can be written as
 \bea \label{general_flux}
 F_{MN}{}^P & = & \mathbb{P}_{({\bf 56+912})M}{}^{a\  R}{}_{b}\tilde{\Omega}_R{}^{b}(t_{a})_N{}^P+\lambda\Omega_{RM}{}^R\delta_N^P\\ \nonumber
 {} & + &   \Omega_{M}{}^0(t_0)_N{}^P-a P_{(adj)}{}^P{}_N{}^R{}_S\Omega_R{}^0(t_0)_M{}^S
 \eea
 or \cite{LeDiffon:2008sh}
 \be
  F_{MN}{}^P= \Theta_M{}^a(t_{a})_N{}^P+(8P_{(adj)}{}^P{}_N{}^R{}_M-\delta_N^P\delta_M^R)\vartheta_R.
 \ee
 We have defined the embedding tensor $\Theta_M{}^a$ as
 \be
 \Theta_M{}^a=7P_{(912)M}{}^{a \ R}{}_{b}\tilde{\Omega}_R{}^{b}
 \ee
 and
 \be
 \vartheta_M=-\frac{1}{2}E_M{}^{\bar{A}}e^{2\Delta}\partial_P(e^{-2\Delta}E_{\bar{A}}{}^P)=-\frac{1}{2}E_M{}^{\bar{A}}\nabla_PE_{\bar{A}}{}^P
 \ee
 is the gauging associated to the trombone symmetry.
 For $E_7\times\mathbb{R}^{+}$
\be
\mathbb{P}_{({\bf 56+912})M}{}^{a\  R}{}_{b}=7P_{(912)M}{}^{a \ R}{}_{b}-\frac{19}{2}P_{(56)M}{}^{a \ R}{}_{b},
\ee
$\lambda=\frac12$ and $a=12$. In fact, taking properly the  projection  $\mathbb{P}_{({\bf R_1+R_3})M}{}^{a\ \ R}{}_{b}$, according to gauged supergravity,  (\ref{general_flux}) is valid for all $E_d\times\mathbb{R}^{+}$ with $d\leqslant7$. In the next section we will  take (\ref{general_flux}) as a conjecture valid for all $d\leqslant11$ and as a first test we will use it to build the generalised transformation for the $E_8\times\mathbb{R}^{+}$ case.

\subsection{The $SL(8)$ decomposition}
The fundamental representation of $E_7$ breaks into the $SL(8)$ group as ${\bf 56 = 28+\overline{28}}$. The index $M$ breaks according to
\be
V^M=(V^{[\hat{\alpha}\hat{\beta}]} \ , \ V_{[\hat{\alpha}\hat{\beta}]})
\ee
where $\hat{\alpha}$ runs from $1$ to $8$, also the components of the generalised Lie derivative (\ref{transformation}) read
\be
({\cal L}_{\xi}V)^M=(({\cal L}_{\xi}V)^{[\hat{\alpha}\hat{\beta}]} \ , \ ({\cal L}_{\xi}V)_{[\hat{\alpha}\hat{\beta}]}).
\ee
The adjoint representation breaks  into the $SL(8)$ group as ${\bf 133 = 63+70}$
\be
V^a=(V_{\hat{\alpha}}{}^{\hat{\beta}} \ , \  V_{[\hat{\alpha}_1\ldots\hat{\alpha}_4]})\qquad;\qquad V_{\ha}{}^{\ha}=0.
\ee
The generators in the $SL(8)$ decomposition \footnote{We take the generalised Kronecker delta defined as $\delta_{\ha_1\ldots\ha_p}^{\hb_1\ldots\hb_p}=\delta_{[\ha_1}^{\hb_1}\ldots\delta_{\ha_p]}^{\hb_p}$ and the alternating tensor $\epsilon_{\ha_1\ldots\ha_p} \epsilon^{\hb_1\ldots\hb_p} = p! \; \delta_{\ha_1\ldots\ha_p}^{\hb_1\ldots\hb_p}$.} take the simple form \cite{LeDiffon:2011wt}
\be
(t_{{\hat\alpha}_1}{}^{{\hat\beta}_1})_{{\hat\alpha}_2{\hat\alpha}_3}{}^{{\hat\beta}_2{\hat\beta}_3}=-\delta_{[{\hat\alpha}_2}^{{\hat\beta}_1}\delta_{{\hat\alpha}_3]
{\hat\alpha}_1}^{{\hat\beta}_2{\hat\beta}_3}
 -\frac18\delta_{{\hat\alpha}_1}^{{\hat\beta}_1}\delta_{{\hat\alpha}_2{\hat\alpha}_3}^{{\hat\beta}_2{\hat\beta}_3}=-(t_{\ha_1}{}^{\hb_1})^{\hb_2\hb_3}{}_{\ha_2\hb_3}
\ee
\bea
(t_{\ha_1\ldots\ha_4})_{\hb_1\ldots\ha_4} & = & \frac{1}{24}\epsilon_{\ha_1\ldots\ha_4\ldots\hb4}\\ \nonumber
(t_{\ha_1\ldots\ha_4})^{\hb_1\ldots\hb_4} & = & \delta_{\ha_1\ldots\ha_4}^{\hb_1\ldots\hb_4}.
\eea
The Cartan-Killing metric is given by
\be
K^{ab}
= \left(\begin{matrix}
 K_{{\hat\alpha}_1}{}^{{\hat\beta}_1}{}_{{\hat\alpha}_2}{}^{{\hat\beta}_2} & 0 \\
0 & \frac{1}{12}\epsilon_{\hat{\alpha}_1\ldots\hat{\alpha}_8}\\
\end{matrix}\right)\label{killing}
\ee
where
\be
 K_{{\hat\alpha}_1}{}^{{\hat\beta}_1}{}_{{\hat\alpha}_2}{}^{{\hat\beta}_2}=3(\delta_{{\hat\alpha}_1}^{{\hat\beta}_2}\delta_{{\hat\alpha}_2}^{{\hat\beta}_1}
 -\frac18\delta_{{\hat\alpha}_1}^{{\hat\beta}_1}\delta_{{\hat\alpha}_2}^{{\hat\beta}_2}).
\ee

Having the decomposition of the Cartan-Killing metric and the generators one can compute the components of (\ref{pro-adj}) involved in (\ref{transformation}),  these are
\bea
P^{\ha_1\ha_2}{}_{\hb_1\hb_2}{}^{\hc_1\hc_2}{}_{\hd_1\hd_2} & = & (t_{\hr_1}{}^{\hs_1})_{\hb_1\hb_2}{}^{\ha_1\ha_2}(K^{-1})^{\hr_1}{}_{\hs_1}{}^{\hr_2}{}_{\hs_2} (t_{\hr_2}{}^{\hs_2})_{\hd_1\hd_2}{}^{\hc_1\hc_2}\\ \nonumber
P^{\ha_1\ha_2\hb_1\hb_2\hc_1\hc_2\hd_1\hd_2} & = & (t_{\hr_1\ldots\hr_4})^{\ha_1\ha_2\hb_1\hb_2}(K^{-1})^{\hr_1\ldots\hr_4\hs_1\ldots\hs_4}(t_{\hs_1\ldots\hs_4})^{\hc_1\hc_2\hd_1\hd_2} \\ \nonumber
P^{\ha_1\ha_2\hb_1\hb_2}{}^{\hc_1\hc_2\hd_1\hd_2} & = & (t_{\hr_1\ldots\hr_4})_{\ha_1\ha_2\hb_1\hb_2}(K^{-1})^{\hr_1\ldots\hr_4\hs_1\ldots\hs_4}(t_{\hs_1\ldots\hs_4})^{\hc_1\hc_2\hd_1\hd_2} \\ \nonumber
P_{\ha_1\ha_2}{}^{\hb_1\hb_2\hc_1\hc_2}{}_{\hd_1\hd_2} & = & -(t_{\hr_1}{}^{\hs_1})_{\ha_1\ha_2}{}^{\hb_1\hb_2}(K^{-1})^{\hr_1}{}_{\hs_1}{}^{\hr_2}{}_{\hs_2} (t_{\hr_2}{}^{\hs_2})_{\hd_1\hd_2}{}^{\hc_1\hc_2}.
\eea
After a long computation the two components of the generalised Lie derivative in the ${\bf 28}$ and $\overline{{\bf 28}}$ representations are given by
\bea\label{deri28}
({\cal L}_{\xi}V)^{\ha_1\ha_2} & = & \xi^{\hr_1\hr_2}\partial_{\hr_1\hr_2}V^{\ha_1\ha_2}+2V^{\hr_1\ha_1}\partial_{\hr_1\hs_1}\xi^{\ha_2\hs_1}
-2V^{\hr_1\ha_2}\partial_{\hr_1\hs_1}\xi^{\ha_1\hs_1}\\ \nonumber
{} & {} & -\frac14\epsilon^{\ha_1\ha_2\hb_1\hb_2\hc_1\hc_2\hd_1\hd_2}V_{\hb_1\hb_2}\partial_{\hc_1\hc_2}\xi_{\hd_1\hd_2}+\partial_{\hr_1\hr_2}\xi^{\hr_1\hr_2}V^{\ha_1\ha_2}
\eea
\bea\label{deribar28}
({\cal L}_{\xi}V)_{\ha_1\ha_2} & = & \xi^{\hr_1\hr_2}\partial_{\hr_1\hr_2}V_{\ha_1\ha_2}+2V_{\hr_1\ha_1}\partial_{\ha_2\hr_2}\xi^{\hr_1\hr_2}
-2V_{\hr_1\ha_2}\partial_{\ha_1\hr_2}\xi^{\hr_1\hr_2}\\ \nonumber
{} & {} & -6V^{\hr_1\hr_2}\partial_{[\hr_1\hr_2}\xi_{\ha_1\ha_2]}.
\eea

The next step in the construction is to look at the $SL(7)$ decomposition and then make the correspondence between the $E_7$ and the seven-dimensional objects, i.e
\bea
V^{\alpha8}=v^{\alpha} &{}& V_{\alpha\beta}=\omega_{\alpha\beta}\\ \nonumber
V^{\alpha\beta}=\frac{1}{5!}\epsilon^{\alpha\beta\gamma_1\ldots\gamma_5}\sigma_{\gamma_1\ldots\gamma_5} &{}&V_{\alpha8}=\frac{1}{7!}\epsilon^{\beta_1\ldots\beta_7}\tau_{\alpha,\beta_1\ldots\beta_7}.
\eea
where the unhatted indices run from $1$ to $7$. Looking at the unhatted components of (\ref{deri28}) and (\ref{deribar28}) we get
\bea\label{transformation_indices}
({\cal L}_{\Vp}\Vpp)^{\alpha} & = & (L_{\vp}\vpp)^{\alpha}\\ \nonumber
({\cal L}_{\Vp}\Vpp)_{\alpha_1\alpha_2} & = & (L_{\vp}\wpp)_{\alpha_1\alpha_2}-(\iota_{\vpp}d\wpr)_{\alpha_1\alpha_2}\\ \nonumber
({\cal L}_{\Vp}\Vpp)_{\alpha_1\ldots\alpha_5} & = & (L_{\vp}\sipp)_{\alpha_1\ldots\alpha_5}-(\iota_{\vpp}d\sip)_{\alpha_1\ldots\alpha_5}-(\wpp\wedge d\wpr)_{\alpha_1\ldots\alpha_5}\\ \nonumber
({\cal L}_{\Vp}\Vpp)_{\alpha,\beta_1\ldots\beta_7} &= &(L_{\vp}\tpp)_{\alpha,\beta_1\ldots\beta_7}-\frac{7!}{1!6!}\wpp_{\alpha[\beta_1}(d\sip)_{\beta_2\ldots\beta_7]}-
\frac{7!}{2!5!}(d\wpr)_{\alpha[\beta_1\beta_2}\sipp_{\beta_3\ldots\beta_7]}.
\eea

Let us  make some remarks about the last two terms of (\ref{transformation_indices}). In order to write the generalised Lie derivative independent of the coordinates \footnote{Notice that the three first lines of (\ref{transformation_indices}) can be straightforwardly written in a coordinate-independent way.} we need to write these two terms in a coordinate-independent way. We note that
\be
\frac{7!}{1!6!}\wpp_{\alpha[\beta_1}(d\sip)_{\beta_2\ldots\beta_7]}=e_{\alpha}{}^{\bar{a}}(\iota_{e_{\bar{a}}}\wpp\wedge d\sip)_{\beta_1\beta_2\ldots\beta_7}
\ee
where $e_{\bar{a}}$ and $e^{\bar{a}}$ are some frames on $TM$ and its dual on $T^{*}M$ respectively.  Defining the function $j$ as
\be
j(\cdot \ ,\ \cdot):\Lambda^nT^*M\otimes\Lambda^pT^*M\rightarrow\Lambda T^*M\otimes\Lambda^{n-1+p}T^*M
\ee
where $(\cdot \ ,\ \cdot)$ has been  explicited just to point out the fact that $j$ is a function with two inputs. This can be written as
\be
j(\cdot \ , \ \cdot)=e^{\bar{a}}(\iota_{e_{\bar{a}}}\ \cdot)\wedge\ \cdot\ ,\label{jota}
\ee
notice that although the frame appears explicitly in the definition it is well defined and is independent of the coordinates.
Collecting the information and plugging it in (\ref{transformation_indices}), the generalised Lie derivative can be written in a coordinate-independent way as
\cite{Coimbra:2011ky}
\bea\label{transformation_no_indices}
({\cal L}_{\Vp}\Vpp)^{1} & = & L_{\vp}\vpp\\ \nonumber
({\cal L}_{\Vp}\Vpp)^{2} & = & L_{\vp}\wpp-\iota_{\vpp}d\wpr\\ \nonumber
({\cal L}_{\Vp}\Vpp)^{3} & = & L_{\vp}\sipp-\iota_{\vpp}d\sip-\wpp\wedge d\wpr\\ \nonumber
({\cal L}_{\Vp}\Vpp)^{4} & = & L_{\vp}\tpp-j(\wpp, d\sip)-j(d\wpr,\sipp).
\eea
Given (\ref{transformation_no_indices}) a natural question that arises is, whether this transformation works beyond seven dimensions. The first we note against a possible lifting is that the last two terms in (\ref{transformation_no_indices}) can be written, in seven dimension, in two equivalent forms. For example,
\be
j(d\wpr, \sipp)=e^{\bar{a}}\iota_{e_{\bar{a}}}d\wpr\wedge\sipp
\ee
and
\be
e^{\bar{a}}\iota_{e_{\bar{a}}}d\wpr\wedge\sipp=e^{\bar{a}}\big(\iota_{e_{\bar{a}}}(d\wpr\wedge\sipp)+d\wpr\wedge\iota_{e_{\bar{a}}}\sipp\big)\label{prueba}
\ee
this implies
\be
j(d\wpr,\sipp)=j(\sipp, d\wpr)\label{doubt}.
\ee
We have been able to do that since the first term of the right hand side of (\ref{prueba}) is identically zero only in seven dimensions. If the manifold had a higher dimension the latter result would be completely different. Consequently, if an extension were possible a reasonable doubt would exist, since we would not be sure which one of both expressions in (\ref{doubt}) is the correct one beyond seven dimensions. The same happens with the other j-terms. In the next section we will see how the Leibniz property for (\ref{transformation_no_indices}) is satisfied only for $n\leqslant7$, giving us another proof that a lifting from seven dimensions is impossible.

\subsection{Consistency conditions}

Consistency conditions of the transformation (\ref{transformation_no_indices}) can be condensed in a single expression like (\ref{algebra}), the antisymmetric part of this expression is called closure of the algebra while the symmetric part is the Leibniz identity. To compute the Leibniz property for (\ref{transformation_no_indices}) we introduce the $\Delta$ operator which provides an elegant way to check the properties of covariance of the generalised objects. It is defined as \cite{Hohm:2011si}, \cite{Aldazabal:2013mya}, \cite{Aldazabal:2013via}
\be
\Delta_{\xi}={\cal L}_{\xi}-\delta_{\xi}
\ee
where the $\delta$ operator is defined through the relation
\be
\delta_{\xi}V={\cal L}_{\xi}V.
\ee
Notice that the latter relation only holds if $V$ is a generalised vector, for example, on a generalised connection $\Gamma$
\be
\delta_{\xi}\Gamma\neq {\cal L}_{\xi}\Gamma,
\ee
also that $\delta_{\xi}(\partial_PV)=\partial_P(\delta_{\xi}V)=\partial_P({\cal L}_{\xi}V)$ but ${\cal L}_{\xi}(\partial_PV)\neq\partial_P({\cal L}_{\xi}V)$.
Using the $\Delta$ operator we have noticed that
\be
\Delta_{V}({\cal L}_{\Vp}\Vpp)=\big[{\cal L}_{V},{\cal L}_{\Vp}\big]\Vpp-{\cal L}_{{\cal L}_{V}\Vp}\Vpp.
\ee
Hence, the generalised Lie derivative $({\cal L}_{\Vp}\Vpp)$ could be identified with a generalised vector only if $({\cal L}_{\Vp}\Vpp)$ transforms property, namely if
\be
\Delta_{V}({\cal L}_{\Vp}\Vpp)=0,
\ee
which actually implies the Leibniz property.

Explicitly we get
\bea\label{algebra1}
(\Delta_{V}({\cal L}_{\Vp}\Vpp))^4 & = & \big(\big[{\cal L}_{V},{\cal L}_{\Vp}\big]\Vpp-{\cal L}_{{\cal L}_{V}\Vp}\Vpp\big)^4  \\ \nonumber
{} & =& e^{\bar{a}}\iota_{\vpp}(\iota_{e_{\bar{a}}}d\omega\wedge d\sip-\iota_{e_{\bar{a}}}d\wpr\wedge d\sigma) -\wpp\wedge d\wpr\wedge d \omega
\eea
the three first components of $\Delta_{V}({\cal L}_{\Vp}\Vpp)$ are identically zero. Notice that in seven dimensions there are no consistency issues since every term in the right hand side of (\ref{algebra1}) is or comes from an $8$-form. What is clear is that, beyond seven dimensions, the Leibniz property does not hold. Let us stress some facts which will be helpful to address the extension beyond seven dimensions. Notice that in (\ref{transformation_no_indices}) the $\iota_{\vpp}$ and $\tp_{(1,7)}$ terms are missing in the fourth component, also that the right hand side of (\ref{algebra1}) is a $(1,7)$ plus an $8$ tensor. These facts are giving us an indication that to achieve the closure of (\ref{algebra1}) for $d>7$, rather than introducing a new parameter lying in a new direction of the generalised tangent space, the transformation (\ref{transformation_no_indices}) has to be completed with the proper $\iota_{\vpp}$ and $\tp_{(1,7)}$ terms \footnote{The $\tp_{(1,7)}$ component for $d>7$ is a $(1,7)+8$ tensor, this will be clarified in the next section.}.

To avoid the inconsistencies in the lifting of (\ref{transformation_no_indices}) to $D=11$ we will move to the $E_8\times\mathbb{R}^{+}$ group. In this case the Leibniz property does not hold from the beginning but, as we will see, there is at least one case where it is possible to move forward to achieve the consistency of the transformation.

\section{$E_8\times\mathbb{R}^{+}$}\label{e8rmas}

\subsection{Summary of previous results}

The $E_8\times\mathbb{R}^{+}$ is trickier since from the beginning one of the two known transformations \cite{Berman:2012vc} does not close and the other \cite{Hohm:2014fxa} needs a new parameter to compensate the failure of the closure of the algebra. The uncomfortable  part of the latter mentioned approach is that this new parameter gives rise to a new degree of freedom which is not present in the $E_{8}\times \mathbb{R}^{+}$ field content, neither in the $E_{11}$ group decomposition. Let us briefly review these two approaches and then present the $SL(9)$ decomposition of the $E_8\times\mathbb{R}^{+}$ generalised Lie derivative.

The proposal of \cite{Berman:2012vc} to the generalised Lie derivative for the $E_8\times\mathbb{R}^{+}$ case  has the same form as in (\ref{transformation}) but with $A^M{}_N{}^P{}_Q$ given by
\be
A^M{}_N{}^P{}_Q=60P_{(adj)}{}^M{}_N{}^P{}_Q-\delta_N^M\delta_Q^P.
\ee
In the $E_8$ group the fundamental and the adjoint representations are essentially the same representation. The generators of the  algebra can be written as
\be
(t^M)_N{}^P=-f^{M}{}_N{}^P\label{structure}
\ee
where $f^{M}{}_N{}^P$ are the structure constants. The generalised vectors are weighted as follows
\be
V^M=e^{-2\Delta}\tilde{V}\qquad;\qquad e^{-2\Delta}=det(e), \label{peso8}
\ee
where $\tilde{V}$ is a pure $E_8$ generalised vector, and the generalised Lie derivative reads
\be
({\cal L}_{\xi}V)^M=\xi^P\partial_PV^M-K_{AB}f^{AM}{}_Nf^{BP}{}_Q\partial_P\xi^QV^N+\partial_P\xi^PV^M.\label{transformation8b}
\ee
This transformation does not satisfy the Leibniz property, its failure is given by
\be
\big(\Delta_{\xi_1}({\cal L}_{\xi_2}V)\big)^M=(\big[{\cal L}_{\xi_1},{\cal L}_{\xi_2}\big]V-{\cal L}_{{\cal L}_{\xi_{1}} \xi_{2}}V)^M=f^{MJ}{}_Nf_I{}^P{}_Q\partial_J\partial_P\xi_{1}^Q\xi_{2}^IV^N.\label{failure}
\ee

The second proposal introduces the parameter $\Sigma$. According to \cite{Hohm:2014fxa}, the generalised transformation is given by
\be
(\hat{\cal L}_{(\xi,\Sigma)}V)^M=\xi^P\partial_PV^M-K_{AB}f^{AM}{}_Nf^{BP}{}_Q\partial_P\xi^QV^N+\partial_P\xi^PV^M-f^{PM}{}_N\Sigma_PV^N.\label{transformation8hs}
\ee
The $\Sigma_P$ parameter is called ``covariant constrained compensating field'' and has to satisfy some constraints. Essentially, these constrains are the same the derivative $\partial_P$  satisfies.
The transformation of the new parameter is fixed by demanding the closure of the algebra, see \cite{Hohm:2014fxa},
\be
\Sigma_{12\,M} =-2\,\Sigma_{[2\,M} \partial_N \Lambda_{1]}^N+2\,\Lambda_{[2}^N \partial_N \Sigma_{1]\,M}
-2\, \Sigma_{[2}^N \partial_M \Lambda_{1]\,N}+f^N{}_{KL} \,\Lambda_{[2}^K\,\partial_M\partial_{N}\Lambda_{1]}^L.\label{lambdaHS}
 \ee

Let us comment about some issues of the transformation (\ref{transformation8hs}). When one defines a generalised Lie derivative, consistency requires it to be independent of the choice of the vector components. As an example \footnote{Particulary, we could take $V=(v,\ \omega,\ 0,\ \tau)$, $\Vp=(\vp,\ \wpr,\ 0,\ \tp)$ and $\Vpp=(\vpp,\ \wpp, 0,\ \tpp)$, hence $\big(\big[{\cal L}_{V},{\cal L}_{\Vp}\big]\Vpp-{\cal L}_{{\cal L}_{V}\Vp}\Vpp\big)^4 =
   -\wpp\wedge d\wpr\wedge d \omega=0$, in $d=7$.}, one can see that in (\ref{transformation_no_indices}) it is possible to turn off whatever components of the generalised vector $\Vp$ and/or $\Vpp$, then check the Leibniz property (\ref{algebra1}), and it will still hold. In this line of thinking, we could  compute the Leibniz property of (\ref{transformation8hs}) with the following choice $(\xi_1,\Sigma_1)=(\xi_1,0)$, $(\xi_2,\Sigma_2)=(\xi_2,0)$ and $(V,\Sigma)=(V,0)$:
\be
\big(\big[\hat{\cal L}_{(\xi_1,0)},\hat{\cal L}_{(\xi_2,0)}\big]-\hat{\cal L}_{\hat{\cal L}_{(\xi_{1},0)}(\xi_{2},0)}\big)(V,0).
\ee
It is straightforward to see that the Leibniz property will fail since
\be
(\hat{\cal L}_{(\xi,0)}V)^M=({\cal L}_{\xi}V)^M,
\ee
notice that the transformation of the right hand side is (\ref{transformation8b}) and this one does not satisfy the Leibniz property.

We want to stress that when one considers the generalised vectors as $(V,\Sigma)$ any generic truncation of the components of these vectors should form a subalgebra of the algebra of the generalised Lie derivative, as happens for all $E_d$-exceptional geometries, with $d\leq7$. The consistency of the transformation holds for all vectors in this generalised tangent space with components $(V,\Sigma)$ and not for a subset of these vectors that excludes the vector $(V,0)$.
This fact gives us a clue that the new parameter $\Sigma$ could not be an independent one. If it were the case, (as we will see, it is not an independent parameter) one of the consequences is that this parameter does not induce a new degree of freedom as in \cite{Hohm:2014fxa}.

\subsection{The $SL(9)$ decomposition}

Before preforming  the $SL(9)$ decomposition of the $E_8\times\mathbb{R}^{+}$ generalised transformation we need to introduce some group-theoretic analysis about the $E_8$ group and its representations \cite{Cartan}, \cite{Godazgar:2013rja}. The fundamental (or adjoint) representation of $\mathfrak{e}_8$  has dimension $248$, thus one may take the generators as the structure constants (\ref{structure}). A vector $V^M$, with $M=1\ldots248$, decomposes into $SL(9)$ as
\be
V^M=(V^{\ha}{}_{\hb},V^{[\ha\hb\hc]},V_{[\ha\hb\hc]})\qquad;\qquad V^{\ha}{}_{\ha}=0
\ee
where the hatted indices run from 1 to 9.
The algebra $\big[t^M,t^N\big]=f^{MN}{}_Pt^P$ of the $E_8$ group in the $SL(9)$ decomposition is:
\begin{gather}\label{e8}
 [ t^{\hat{\alpha}}{}_{\hat{\beta}}, t^{\hat{\gamma}}{}_{\hat{\delta}} ] = \delta^{\hat{\gamma}}_{\hat{\beta}} t^{\hat{\alpha}}{}_{\hat{\delta}}
- \delta^{\hat{\alpha}}_{\hat{\delta}} t^{\hat{\gamma}}{}_{\hat{\beta}} \\ \nonumber
[t^{\hat{\alpha}}{}_{\hat{\beta}}, t^{ \hat{\gamma}_{1} \dots \hat{\gamma}_{3}}] = 3 \delta_{\hat{\beta}}^{[\hat{\gamma}_{1}} t^{ \hat{\gamma}_{2} \hat{\gamma}_{3}] \hat{\alpha}} -\frac{1}{3} \delta^{\hat{\alpha}}_{\hat{\beta}} t^{\hat{\gamma}_{1} \dots \hat{\gamma}_{3}}, \\ \nonumber
[t^{\hat{\alpha}}{}_{\hat{\beta}}, t_{ \hat{\gamma}_{1} \dots \hat{\gamma}_{3}}] = - 3 \delta^{\hat{\alpha}}_{[\hat{\gamma}_{1}} t_{ \hat{\gamma}_{2} \hat{\gamma}_{3}] \hat{\beta}} +\frac{1}{3} \delta^{\hat{\alpha}}_{\hat{\beta}} t_{\hat{\gamma}_{1} \dots \hat{\gamma}_{3}},\\ \nonumber
[t^{\hat{\alpha}_{1} \dots \hat{\alpha}_3}, t_{\hat{\beta}_1 \dots \hat{\beta}_3}] = 18 \delta^{[\hat{\alpha}_1 \hat{\alpha}_2}_{[\hat{\beta}_1 \hat{\beta}_2} t^{\hat{\alpha}_3]}{}_{\hat{\beta}_3]}, \\ \nonumber
[t^{\hat{\alpha}_1 \dots \hat{\alpha}_3}, t^{\hat{\beta}_1 \dots \hat{\beta}_3}] = - \frac{1}{3!} \epsilon^{ \hat{\alpha}_1 \hat{\alpha}_2 \hat{\alpha}_3 \hat{\beta}_1 \hat
{\beta}_2 \hat{\beta}_3 \hat{\gamma}_1 \hat{\gamma}_2 \hat{\gamma}_3 } t_{ \hat{\gamma}_1 \dots \hat{\gamma}_3},\\ \nonumber
[t_{\hat{\alpha}_1 \dots \hat{\alpha}_3}, t_{\hat{\beta}_1 \dots \hat{\beta}_3}] = \frac{1}{3!} \epsilon_{ \hat{\alpha}_1 \hat{\alpha}_2 \hat{\alpha}_3 \hat{\beta}_1 \hat{\beta}_2 \hat{\beta}_3 \hat{\gamma}_1 \hat{\gamma}_2 \hat{\gamma}_3 } t^{ \hat{\gamma}_1 \dots \hat{\gamma}_3}.\nonumber
\end{gather}
From (\ref{e8}) it is possible to read off the structure constants:
\bea\label{estructure}
f^{\ha_1}{}_{\hb_1}{}^{\ha_2}{}_{\hb_2}{}_{\ha_3}{}^{\hb_3} & = &\delta_{\hb_1}^{\ha_2}\delta_{\ha_3}^{\ha_1}\delta_{\hb_2}^{\hb_3}-\delta_{\hb_2}^{\ha_1}
\delta_{\ha_3}^{\ha_2}\delta_{\hb_1}^{\hb_3},\\ \nonumber
f^{\ha_1}{}_{\hb_1}{}^{\ha_2\hb_2\hc_3}{}_{\ha_3\hb_3\hc_3} & = & 3\delta_{\hb_1}^{[\ha_2}\delta_{\ha_3\hb_3\hc_3}^{\hb_2\hc_2]\ha_1}-\frac13
\delta_{\hb_1}^{\ha_1}\delta_{\ha_3\hb_3\hc_3}^{\ha_2\hb_2\hc_2},\\ \nonumber
f^{\ha_1}{}_{\hb_1}{}_{\ha_2\hb_2\hc_3}{}^{\ha_3\hb_3\hc_3} & = & -3\delta_{[\ha_2}^{\ha_1}\delta^{\ha_3\hb_3\hc_3}_{\hb_2\hc_2]\hb_1}+\frac13
\delta_{\hb_1}^{\ha_1}\delta^{\ha_3\hb_3\hc_3}_{\ha_2\hb_2\hc_2},\\ \nonumber
f^{\ha_1\hb_1\hc_1}{}_{\ha_2\hb_2\hc_2}{}_{\ha_3}{}^{\hb3} & = & 18\delta_{[\ha_2}^{\hb_3}\delta_{\hb_2\hc_2]}^{[\ha_1\hb_1}\delta_{\ha_3}^{\hc_1]}
,\\ \nonumber
f^{\ha_1\hb_1\hc_1\ha_2\hb_2\hc_2\ha_3\hb_3\hc_3} & = & -\frac{1}{3!}\epsilon^{\ha_1\hb_1\hc_1\ha_2\hb_2\hc_2\ha_3\hb_3\hc_3},\\ \nonumber
f_{\ha_1\hb_1\hc_1\ha_2\hb_2\hc_2\ha_3\hb_3\hc_3} & = & \frac{1}{3!}\epsilon_{\ha_1\hb_1\hc_1\ha_2\hb_2\hc_2\ha_3\hb_3\hc_3}.
\eea

Having the structure constants we are ready to compute the Cartan-Killing
 metric, defined as
\be
K^{AB}=\frac{1}{60}f^{AK}{}_Lf^{BL}{}_K.\label{cartan}
\ee
After a very long computation we get
\be
K^{AB}
= \left(\begin{matrix}
 K_{{\hat\alpha}_1}{}^{{\hat\beta}_1}{}_{{\hat\alpha}_2}{}^{{\hat\beta}_2} & 0 & 0 \\
0 & 0 & 6\delta_{\ha_2\hb_2\hc_2}^{{\ha_1\hb_1\hc_1}}\\
0 & 6\delta_{\ha_1\hb_1\hc_1}^{{\ha_2\hb_2\hc_2}} & 0
\end{matrix}\right)\label{killing8}
\ee
where
\be
 K_{{\hat\alpha}_1}{}^{{\hat\beta}_1}{}_{{\hat\alpha}_2}{}^{{\hat\beta}_2}=\delta_{{\hat\alpha}_1}^{{\hat\beta}_2}\delta_{{\hat\alpha}_2}^{{\hat\beta}_1}
 -\frac19\delta_{{\hat\alpha}_1}^{{\hat\beta}_1}\delta_{{\hat\alpha}_2}^{{\hat\beta}_2}
\ee
and the identity takes the simple form
\be
\delta^{B}_A
= \left(\begin{matrix}\label{ident}
 K_{{\hat\alpha}_1}{}^{{\hat\beta}_1}{}_{{\hat\alpha}_2}{}^{{\hat\beta}_2} & 0 & 0 \\
0 & \delta_{\ha_2\hb_2\hc_2}^{{\ha_1\hb_1\hc_1}} & 0\\
0 & 0 & \delta_{\ha_1\hb_1\hc_1}^{{\ha_2\hb_2\hc_2}}
\end{matrix}\right).
\ee
Notice that $\delta_A^A=248$.

To have a better idea about what is going on with the transformation of the $E_8\times\mathbb{R}^{+}$ group we follow the discussion performing the $SL(9)$ split of (\ref{transformation8b}), but first we will show how the derivative $\partial_{\alpha}$ can be viewed as a section of the dual generalised tangent bundle.
The partial derivative breaks according to
\be
\partial_M=(\partial_{\ha}{}^{\hb},\partial_{\ha\hb\hc},\partial^{\ha\hb\hc})=(\partial_{\ha}{}^{\hb},0,0)\label{emb1}
\ee
and
\be
\partial_{\ha}{}^{\hb}\ \ \ \rightarrow\ \ \ \partial_{\alpha}{}^{9}=\partial_{\alpha}\ \ ,\ \ \partial_{9}{}^{\beta}=\partial_{\alpha}{}^{\beta}=0.\label{emb2}
\ee
This embedding is a solution of the equations (section condition) \cite{Berman:2012vc}, \cite{Hohm:2014fxa}
\bea\label{seccon8}
K^{MN}\partial_M\otimes\partial_N & = & 0\\ \nonumber
f^{AMN}\partial_M\otimes\partial_N & = & 0\\ \nonumber
(f^{A(M}{}_Pf_A{}^{N)}{}_Q-2\delta_P^{(M}\delta_Q^{N)})\partial_M\otimes\partial_N & = & 0
\eea
which implies
\be
Y^M{}_N{}^P{}_Q\partial_M\otimes\partial_P=0,
\ee
where the $Y$ tensor is defined as in (\ref{Y}), but now adapted to the $E_8\times\mathbb{R}^{+}$ case.

A very tedious calculation leads to the three components of the generalised transformation of the $E_8\times\mathbb{R}^{+}$, the first one is given by
\bea\label{Lie1}
({\cal L}_{\xi}V)^{\ha}{}_{\hb} & = & \xi^{\alpha_1}{}_9\partial_{\alpha_1}V^{\ha}{}_{\hb}-V^{\alpha_1}{}_{\hb}\partial_{\alpha_1}\xi^{\ha}{}_9+\partial_{\alpha_1}{}^{\ha}
\xi^{\alpha_1}{}_{\hb_1}V^{\hb_1}{}_{\hb}\\ \nonumber
{} & {} & +\partial_{\hb}\xi^{\hb_1}{}_9V^{\ha}{}_{\hb_1}-\partial_{\alpha_1}\xi^{\alpha_1}{}_{\hb}V^{\ha}{}_{9}+\partial_{\ha_1}\xi^{\ha_1}V^{\ha}{}_{\hb} \\ \nonumber
{} & {} & +\frac96\delta_{[\ha_1}^{\ha}V_{\hb_1\hc_1]\hb}\partial_{\ha_2}{}^{[\ha_1}\xi^{\hb_1\hc_1]\ha_2}-\frac16
\delta^{\ha}_{\hb}\partial_{\ha_2}{}^{[\ha_1}\xi^{\hb_1\hc_1]\ha_2}V_{\ha_1\hb_1\hc_1}\\
\nonumber
{} & {} & +\frac96\partial_{[\ha_1}{}^{\ha_2}\xi_{\hb_1\hc_1]\ha_2}\delta_{\hb}^{[\ha_1}V^{\hb_1\hc_1]\ha}-\frac16
\delta^{\ha}_{\hb}\partial_{[\ha_1}{}^{\ha_2}\xi_{\hb_1\hc_1]\ha_2}V^{\ha_1\hb_1\hc_1},
\eea
and as expected
\be
({\cal L}_{\xi}V)^{\ha}{}_{\ha}=0.
\ee
The other two components are given by
\bea
({\cal L}_{\xi}V)^{\ha_1\hb_1\hc_1} & = & \xi^{\alpha}\partial_{\alpha}V^{\ha_1\hb_1\hc_1}-3\partial_{\alpha}\xi^{[\ha_1}V^{\hb_1\hc_1]\alpha}+
\partial_{\alpha}\xi^{\alpha}V^{\ha_1\hb_1\hc_1}\\ \nonumber
{} & {} & -9V^{\ha_3}{}_{[\ha_2}\delta_{\hb_2\hc_2]\ha_3}^{\ha_1\hb_1\hc_1}\partial_{\alpha}{}^{[\ha_2}\xi^{\hb_2\hc_2]\alpha}\\ \nonumber
{} & {} & +\frac{1}{12}\epsilon^{\ha_1\hb_1\hc_1\ha_2\hb_2\hc_2\ha_3\hb_3\hc_3}\partial_{[\ha_2}\xi_{\hb_2\hc_2]9}V_{\ha_3\hb_3\hc_3}
\eea
and
\bea\label{Lie3}
({\cal L}_{\xi}V)_{\ha_1\hb_1\hc_1} & = & \xi^{\alpha}\partial_{\alpha}V_{\ha_1\hb_1\hc_1}+3V_{\ha[\hb_1\hc_1}\partial_{\ha_1]}\xi^{\ha}-3\partial_{\alpha}\xi^{\alpha}{}_{[\ha_1}V_{\hb_1\hc_1]9}+
\partial_{\alpha}\xi^{\alpha}V_{\ha_1\hb_1\hc_1}\\ \nonumber
{} & {} & -3\big(V^{\ha}{}_{\ha_1}\partial_{[\hb_1}\xi_{\hc_1\ha]9}
          + V^{\ha}{}_{\hc_1}\partial_{[\ha_1}\xi_{\hb_1\ha]9}
          +V^{\ha}{}_{\hb_1}\partial_{[\hc_1}\xi_{\ha_1\ha]9}\big)\\ \nonumber
{} & {} & +\frac{1}{12}\epsilon_{9\hb_2\hc_2\ha_1\hb_1\hc_1\ha_3\hb_3\hc_3}\partial_{\alpha}\xi^{\hb_2\hc_2\alpha}V^{\ha_3\hb_3\hc_3},
\eea
notice that the indices $\ha_1\ ,\hb_1$ and $\hc_1$ in the parenthesis of the expression (\ref{Lie3}) are fully antisymmetrised.

The next step, as in the $E_7\times\mathbb{R}^{+}$ case, is to look at only the unhatted components ($SL(9)\rightarrow SL(8)$) and then associating them with their corresponding eight-dimensional objects. Associating the components of an $E_8$ generalised vector in the $SL(9)$ representation and the components in the $SL(8)$ decomposition is more difficult than for the $E_7$ group since the fundamental representation of $\mathfrak{e}_8$ has dimension $248$.
\begin{table}[!ht]
\begin{center}
\begin{tabular}{c c}
  $V^M$  & $SL(8)$ repr\\
  $V^{\alpha}{}_9 = v^{\alpha}$ & 8\\
  $V_{\alpha_1\alpha_29}=\omega_{\alpha_1\alpha_2}$ & 28 \\
  $V^{\beta_1\beta_2\beta_3}=-\frac{1}{5!}\epsilon^{9\beta_1\beta_2\beta_3\alpha_1\ldots\alpha_5}\sigma_{\alpha_1\ldots\alpha_5}$ & 56\\
  $V^{\alpha}{}_{\beta}=\frac{1}{7!}\epsilon^{9\alpha\gamma_1\ldots\gamma_7}\big(\tau_{\beta,\gamma_1\ldots\gamma_7}+\Lambda_{\beta\gamma_1\ldots\gamma_7}\big)$ & 63+1 \\
  $V_{\beta_1\beta_2\beta_3}=\frac{1}{8!}\epsilon^{9\gamma_1\ldots\gamma_8}\xi_{\beta_1\beta_2\beta_3,\gamma_1\ldots\gamma_8}$ & 56\\
  $V^{\beta_1\beta_29}=\frac{1}{6!8!}\epsilon^{9\beta_1\beta_2\alpha_1\ldots\alpha_6}\epsilon^{9\gamma_1\ldots\gamma_8}
  \xi_{\alpha_1\ldots\alpha_6,\gamma_1\ldots\gamma_8}$ & 28 \\
  $V^9{}_{\gamma}=\frac{1}{8!8!}\epsilon^{9\alpha_1\ldots\alpha_8}\epsilon^{9\beta_1\ldots\beta_8}\xi_{\gamma,\alpha_1\ldots\alpha_8,
  \beta_1\ldots\beta_8}$ & 8\\
\end{tabular}
\end{center}
\caption{\small{$SL(8)$} decomposition of $E_8$}\label{tabla}
\end{table}
From table \ref{tabla} it is possible to see the relation between these two representations, also that the generalised tangent bundle, locally, is
\bea\label{tangent8}
E\simeq TM\oplus \Lambda^2T^*M\oplus \Lambda^5T^*M\oplus (\Lambda T^*M\otimes\Lambda^7T^*M)\\ \nonumber
\oplus(\Lambda^3 T^*M\otimes\Lambda^8T^*M)\oplus(\Lambda^6 T^*M\otimes\Lambda^8T^*M)\oplus(\Lambda T^*M\otimes(\Lambda^8 T^*M)^2),
\eea
see \cite{Strickland-Constable:2013xta} for a discussion about this. Now a generalised vector is represented in components \footnote{Notice that $\tau_{[\beta,\gamma_1\ldots\gamma_7]=0}$.} as
\be
V=(v\ , \  \omega_2 \ , \ \sigma_5 \ , \ \tau_{(1,7)}+\Lambda_{(8)} \ , \ \xi_{(3,8)} \ , \ \xi_{(6,8)}  \ , \ \xi_{(1,8,8)}).\label{vector8}
\ee

As a first check, we will write down explicitly, in terms of the $SL(8)$ fields, the $E_8\times\mathbb{R}^{+}$ transformation up to the dual diffeomorphism. This is the reason we will present only the components of the generalised Lie derivative that correspond with the first line of (\ref{tangent8}), or the four first lines of table \ref{tabla}.
The local expressions of the remaining components of the generalised Lie derivative are quite complicated ones and highly non covariant. We consider that they will give no relevant information for the subsequent analysis.
However, we want to stress that despite we will not present this components, we are not ignoring the other $92$ generators since they are fully considered in (\ref{Lie1})-(\ref{Lie3}) which is the full expression, so far, of the generalised Lie derivative.

The three first components $({\cal L}_{\xi}V)^{\alpha}{}_9$, $({\cal L}_{\xi}V)_{\alpha_1\alpha_29}$ and $({\cal L}_{\xi}V)^{\alpha_1\alpha_2\alpha_3}$ can be straightforwardly computed, getting
\bea\label{88}
({\cal L}_{\Vp}\Vpp)^{\alpha} & = & (L_{\vp}\vpp)^{\alpha}\\ \nonumber
({\cal L}_{\Vp}\Vpp)_{\alpha_1\alpha_2} & = & (L_{\vp}\wpp)_{\alpha_1\alpha_2}-(\iota_{\vpp}d\wpr)_{\alpha_1\alpha_2}\\ \nonumber
({\cal L}_{\Vp}\Vpp)_{\alpha_1\ldots\alpha_5} & = & (L_{\vp}\sipp)_{\alpha_1\ldots\alpha_5}-(\iota_{\vpp}d\sip)_{\alpha_1\ldots\alpha_5}-(\wpp\wedge d\wpr)_{\alpha_1\ldots\alpha_5}.
\eea
As in the $E_7\times\mathbb{R}^{+}$ case these three components can be written in a coordinate-independent way.
The $({\cal L} V)^{\alpha}{}_{\beta}$ component is quite hard to compute but at the end we get
\bea\nonumber\label{dualdif}
({\cal L}_{\Vp} \Vpp)^{\alpha}{}_{\beta} & = & \frac{1}{7!}\epsilon^{\alpha\gamma_1\ldots\gamma_7}\Big((L_{\vp}\tpp)_{\beta,\gamma_1\ldots\gamma_7}
-8v^{\prime\prime\rho}(\partial_{\underline{\rho}}\tp_{\beta,\underline{\gamma_1\ldots\gamma_7}}+\partial_{\beta}\Lambda^{\prime}_{\rho\gamma_1\ldots\gamma_7}
-\partial_{\underline{\beta}}\Lambda^{\prime}_{\rho\underline{\gamma_1\ldots\gamma_7}})\\ \nonumber
{} & {} & -j(\wpp ,d\sip)_{\beta,\gamma_1\ldots\gamma_7}+\frac14(\wpp\wedge d\sip)_{\beta\gamma_1\ldots\gamma_7} \\ \nonumber
{} & {} & -j(d\wpr ,\sipp)_{\beta,\gamma_1\ldots\gamma_7}+\frac38(d\wpr\wedge\sipp)_{\beta\gamma_1\ldots\gamma_7}\Big)\\ \nonumber
{} & {} & +\delta^{\alpha}_{\beta}\Big(L_{\vp}\Lambda^{\prime\prime}-\frac{8}{8!}v^{\prime\prime\rho}(\partial_{\underline{\alpha_1}}
\Lambda^{\prime}_{\rho\underline{\gamma_1\ldots\gamma_7}}+\partial_{\underline{\rho}}\Lambda^{\prime}_{\alpha_1\underline{\gamma_1\ldots\gamma_7}}
)\epsilon^{\alpha_1\gamma_1\ldots\gamma_7}\\
{} & {} & +\frac{1}{12}\wpp\wedge d\sip-\frac{1}{24} d\wpr\wedge\sipp
\Big),
\eea
where the underline indices are fully antisymmetrised.

Let us point out some facts about (\ref{dualdif}). First, we have explicitly separated the reducible representation ${\bf 64}\rightarrow{\bf 63+1}\rightarrow {\cal L}\tau+{\cal L}\Lambda$. Notice that  if $d=7$ we recover the expression of the fourth component of the generalised Lie derivative (\ref{transformation_no_indices}). Ignoring for the moment the $\Lambda$ transformation we have
\bea\nonumber\label{dualdif1}
({\cal L}_{\Vp} \Vpp)_{\beta,\gamma_1\ldots\gamma_7}  & = & (L_{\vp}\tpp)_{\beta,\gamma_1\ldots\gamma_7}
-8v^{\prime\prime\rho}(\partial_{\underline{\rho}}\tp_{\beta,\underline{\gamma_1\ldots\gamma_7}}+\partial_{\beta}\Lambda^{\prime}_{\rho\gamma_1\ldots\gamma_7}
-\partial_{\underline{\beta}}\Lambda^{\prime}_{\rho\underline{\gamma_1\ldots\gamma_7}})\\ \nonumber
{} & {} & -j(\wpp ,d\sip)_{\beta,\gamma_1\ldots\gamma_7}+\frac14(\wpp\wedge d\sip)_{\beta\gamma_1\ldots\gamma_7} \\
{} & {} & -j(d\wpr ,\sipp)_{\beta,\gamma_1\ldots\gamma_7}+\frac38(d\wpr\wedge\sipp)_{\beta\gamma_1\ldots\gamma_7}.
\eea
Now, it is easier to check  $({\cal L}_{\Vp} \Vpp)_{[\alpha,\gamma_1\ldots\gamma_7]}=0$. On the other hand one may see that getting a coordinate-independent writing of (\ref{dualdif1}) is impossible since the equivalent term to $-(\iota_{\vpp}d)$ in (\ref{dualdif1}) is
\be
-8v^{\prime\prime\rho}(\partial_{\underline{\rho}}\tp_{\beta,\underline{\gamma_1\ldots\gamma_7}}+\partial_{\beta}\Lambda^{\prime}_{\rho\gamma_1\ldots\gamma_7}
-\partial_{\underline{\beta}}\Lambda^{\prime}_{\rho\underline{\gamma_1\ldots\gamma_7}})\label{curt}
\ee
and this term  has a coordinate-dependent writing, namely, this term is not a tensor. It is one of the reasons why the covariance can not be achieved in the $E_8$ generalised Lie derivative. Notice that the terms in the parenthesis of (\ref{curt}) are exactly the transformation predicted  \cite{Curtright:1980yk}, \cite{Aulakh:1986cb}  for a tensor with a $(1,8)$ mixed symmetry, on a linearised background.

 This is not the end of the story, in the next section we will see that as long as the theory is defined on a generalised parallelisable manifold the transformation
can be consistently defined in $d=8$ and perhaps extended to $d>8$.

\subsection{Building the $E_8\times\mathbb{R}^{+}$ generalised transformation}\label{building}

To built the $E_8\times\mathbb{R}^{+}$ generalised transformation the starting point will be (\ref{general_flux}), but adapted to this case. The most general theory of gauged supergravity in $d=3$, where the trombone symmetry is gauged \cite{LeDiffon:2008sh}, needs, for consistency,  gaugings living in the ${\bf 1+248+3875}$ representations of the $E_8$ group. Taking that into account, the proposal, for the $E_8$ fluxes, is as follows
\bea \label{general_flux8}
 F_{MN}{}^P & = & \mathbb{P}_{({\bf 1+248+3875})M}{}^{A\ \ R}{}_{B}\tilde{\Omega}_R{}^{B}(t_{A})_N{}^P+\lambda\Omega_{RM}{}^R\delta_N^P\\ \nonumber
 {} & {} &  + \Omega_{M}{}^0(t_0)_N{}^P-a P_{(adj)}{}^P{}_N{}^R{}_S\Omega_R{}^0(t_0)_M{}^S,
 \eea
 where, given (\ref{peso8}),
 \be
 \Omega_{M}{}^0=2\partial_M\Delta,
 \ee
 notice that all indices are denoted by upper case. To fix $\mathbb{P}_{({\bf 1+248+3875})M}{}^{A\ \ R}{}_{B}$ we need to know explicitly the projectors to each representation, these are given by \cite{LeDiffon:2008sh}
\bea
P_{(1)M}{}^{NP}{}_Q & = & \frac{1}{248}\delta_M^N\delta_Q^P\\ \nonumber
P_{(248)M}{}^{NP}{}_Q & = &\frac{1}{60}f^{AN}{}_Mf_A{}^P{}_Q \\ \nonumber
P_{(3875)M}{}^{NP}{}_Q & = &\frac{1}{28}f^{AP}{}_Mf_A{}^N{}_Q-\frac{1}{28}f^{ANP}f_{AMQ}+\frac{1}{14}\delta_M^P\delta_Q^N-\frac{1}{56}\delta_M^N\delta_Q^P+\frac{1}{14}K^{PN}K_{QM}.
\eea
Then $\mathbb{P}_{({\bf 1+248+3875})}$ can be written as a linear combination of these projectors, i.e
\be
\mathbb{P}_{({\bf 1+248+3875})}=a_1P_{(1)}+a_2P_{(248)}+a_3P_{(3875)}.
\ee
The coefficients $a_i$ can be fixed demanding $ F_{MN}{}^P$ (or in planar indices $F_{\bar{A}\bar{B}}{}^{\bar{C}}$) is the generalised transformation \footnote{We know (\ref{transformation8b}) is not a well defined transformation, but from the $SL(9)$ decomposition (\ref{88}) and (\ref{dualdif}) we already know (\ref{transformation8b}) is close to the right one.} (\ref{transformation8b}) with all vectors being frames, i.e  $ (\hat{\cal L}_{E_{\bar{A}}} E_{\bar{B}})^{M}= F_{\bar{A}\bar{B}}{}^{\bar{C}}E_{\bar{C}}{}^{M} $.
We see that taking $a_1=62$, $a_2=-30$, $a_3=14$, $\lambda=1$ and $a=60$ the fluxes can be written as
\bea\label{flujoderi}
F_{\bar{A}\bar{B}}{}^{\bar{C}}E_{\bar{C}}{}^{M} & = & E_{\bar{A}}{}^{P}\partial_{P}E_{\bar{B}}{}^{M}-f_{A}{}^M{}_Nf^{AP}{}_Q\partial_P
E_{\bar{A}}{}^{Q}E_{\bar{B}}{}^{N}+\partial_{P}E_{\bar{A}}{}^{P}E_{\bar{B}}{}^{M}\\ \nonumber
{} & {} & +K_{IQ}\tilde{\Omega}_P{}^Q(t^P)_J{}^ME_{\bar{A}}{}^{I}E_{\bar{B}}{}^{J}.
\eea
From the right hand side of (\ref{flujoderi}) it is possible to read off the generalised Lie derivative.

The next step is to extend the transformation to general vectors and not only on frames, namely $E_{\bar{A}}\rightarrow\xi$ and $E_{\bar{B}}\rightarrow V$, getting
\be
(\hat{{\cal L}}_{\xi}V)^M=\xi^P\partial_PV^M-f_A{}^{M}{}_Nf^{AP}{}_Q\partial_P\xi^QV^N+\partial_P\xi^PV^M-f^{MP}{}_Q\Sigma_PV^Q.\label{transformationyo}
\ee
It is remarkable that we have obtained a generalised Lie derivative that has the same form as the one presented in \cite{Hohm:2014fxa} but in our transformation $\Sigma$ is not a parameter, this can be written by means of a connection as
\be
\Sigma_P=\frac{1}{60}f_J{}^K{}_L\tilde{\Omega}_{PK}{}^L\xi^J\label{effecpara}
\ee
where we have used $\tilde{\Omega}_{PK}{}^L=\tilde{\Omega}_{P}{}^Q(t_{Q})_K{}^L$, we recall
$\tilde{\Omega}_{PK}{}^L=\tilde{E}_K{}^{\bar{A}}\partial_P\tilde{E}_{\bar{A}}{}^L$.
 Furthermore, the index $P$ is one that corresponds to a derivative, hence it is straightforward to see that under the section condition $\Sigma_P$ behaves
as the derivative $\partial_{P}$ does.
Also, due to the fact that $f_{J}{}^K{}_L$ and $\tilde{\Omega}_{PK}{}^L$ have weights $\lambda=0$ and $\lambda=1$ respectively, $\Sigma_P$ has zero weight.
Moreover, one can see that this is the only transformation that can be built without introducing further parameters or degrees of freedom on the manifold. Also, by construction (\ref{transformationyo}) is consistent with gauged supergravity in three dimensions. Regarding this statement, now we are able to write the generalised fluxes (\ref{general_flux8}) as in \cite{LeDiffon:2008sh},
\be
F_{MN}{}^P=\Theta_M{}^A(t_A)_N{}^P+\big(\frac{1}{2}(t^A)_N{}^P(t_A)_M{}^Q-\delta_N^P\delta_M^Q\big)\vartheta_Q
\ee
 where
 \be
 \Theta_M{}^A=\mathbb{P}_{({\bf 1+3875})M}{}^{A\ \ R}{}_{B}\tilde{\Omega}_R{}^{B}
 \ee
 is the embedding tensor in the ${\bf 1+3875}$ representations of the $E_8$ group, and the gauging associated to the trombone symmetry is written as
 \be
 \vartheta_M=-E_M{}^{\bar{A}}e^{2\Delta}\partial_P(e^{-2\Delta}E_{\bar{A}}{}^P)=-E_M{}^{\bar{A}}\nabla_PE_{\bar{A}}{}^P,
 \ee
 we recall that $E_{\bar{A}}{}^P=e^{-2\Delta}\tilde{E}_{\bar{A}}{}^P$. In general, provided $\Theta_M{}^A$ and $\vartheta_Q$ are not constants in this context, $F_{MN}{}^P$ is not a constant.

In order to write the generalised transformation we have introduced the generalised Weitzenb\"{o}ck connection which is defined in terms of the generalised frame as in (\ref{weitzenbock}). the generalised frame encodes the degrees of freedom of the theory, \footnote{In general the frame is the degree of freedom of the theory but there is one special frame which can be written in terms of $e^{\bar{a}}$, $A_3$, $A_6$, $A_{(1,8)}$, $A_{(3,9)}$, $A_{(6,9)}$, $A_{(1,8,9)}$.} for this reason, no further fields or parameters are needed to get a well defined transformation. However, something interesting is happening, since now the parameters and the degrees of freedom will be mixed in the transformation, as happens in closed string field theory (CSFT), see  \cite{Zwiebach:1992ie} for a discussion on gauge transformations in CSFT.

Let us, conjectural and schematically, discuss this statement in this context. DFT or GG, for instance, can be seen as a consistent truncation of CSFT \cite{Hull:2009mi}, \cite{Hull:2009zb}, \cite{Hohm:2014xsa}. This truncation is performed in such a way that the truncated algebra of CSFT closes off-shell and the truncated gauge parameters transform in the fundamental representation of the $O(d,d)$ group. Hence, the parameters can be regarded as generalised vectors transforming under the action of the $O(d,d)$ group.

In the exceptional generalised geometries one expects that something similar takes place, and it would be desirable to identify these geometries with a consistent truncation of some string field theory (SFT). In fact, in all exceptional geometries for $d\leq7$ the parameters (generalised vectors) transform only in the fundamental representation of the exceptional group. In $d=8$, from the generalised Lie derivative (\ref{transformationyo}), one can see that the generalised vectors also transform restricted to the fundamental representation which at the same time is the adjoint representation of the $E_8$ group \footnote{We recall that in generalised geometry the parameters (generalised vectors) belong to the fundamental representation of the group, while the degrees of freedom belong to the adjoint one.}. This means that, in the same way that the field degrees of freedom need the parameters to transform, the parameters, in the $E_8$ case, need the field degrees of freedom to transform.

In  CSFT usually the algebra of the gauge parameters is represented as follow
\be
[\delta_{\xi_1},\delta_{\xi_2}]|\Psi\rangle=\delta_{\xi_{12}(\Psi)}|\Psi\rangle\ +\ (\text{on-shell=0 terms})\label{CSFT}
\ee
where $|\Psi\rangle$ is the string field containing the excitations of string and, it is said, $\xi_{12}(\Psi)$ is a field dependent parameter.
Translating it to our case, the equivalent to the truncated string field should be the generalised frame $E_{\bar{A}}$, strictly speaking, it should be a combination of certain components of the frame. The equivalent to the parameter  $\xi_{12}(\Psi)$, in our language, should be the generalised transformation $\hat{\delta}_{\xi_1}\xi_2={\hat{\cal L}}_{\xi_1}\xi_2$, (\ref{transformationyo}), which is a field dependent one. The equivalent $E_8\times\mathbb{R}^{+}$ expression to (\ref{CSFT}) should be
 \be
[\hat{\cal L}_{\xi_1},\hat{\cal L}_{\xi_2}]E_{\bar{A}}=\hat{\cal L}_{\hat{\cal L}_{\xi_{1}}\xi_2}E_{\bar{A}},
\ee
which is the Leibniz property of our theory. These facts are showing an intriguing analogy between some hypothetical SFT (truncated) and the exceptional approach,  beyond the consistent truncation of DFT \cite{Hull:2009mi}, \cite{Hull:2009zb}, \cite{Hohm:2014xsa},  which is worth exploring, and maybe relate this SFT  with some theory based on the $E_{11}$ group \cite{West:2001as}.

To know the full form of the fourth component of the $E_8\times\mathbb{R}^{+}$ generalised transformation, we need to know the $SL(8)$ decomposition of $f^{MP}{}_Q\Sigma_PV^Q$. Taking into account
\be
\Sigma_P=(\Sigma_{\hb}{}^{\ha},\Sigma_{\ha\hb\hc},\Sigma^{\ha\hb\hc})=(\Sigma_{\hb}{}^{\ha},0,0)\label{embeding},
\ee
where
\be
\Sigma_{\hb}{}^{\ha}\ \ \ \rightarrow\ \ \ \Sigma_{\beta}{}^{9}=\Sigma_{\beta}\ \ ,\ \ \Sigma_{9}{}^{\alpha}=\Sigma_{\beta}{}^{\alpha}=0.
\ee
and (\ref{estructure})
we get
\be
f^{MP}{}_Q\Sigma_PV^Q=\Big((\Sigma_{\beta}v^{\alpha}-\frac18\delta_{\beta}^{\alpha}\Sigma_{\rho}v^{\rho})+\frac18\delta_{\beta}^{\alpha}
\Sigma_{\rho}v^{\rho}
\ ,0 \ ,0\Big)\label{sigmadeco}.
\ee
The final expression for the ${\bf 63+1}$ component can be read from (\ref{dualdif}) plus the terms (\ref{sigmadeco}), with $\Sigma_{\beta}=\frac{1}{60}f_J{}^K{}_L\tilde{\Omega}_{\beta K}{}^L V^{\prime J}$.

\subsection{Consistency conditions and compatibility}
\subsection*{Leibniz property}
Now we proceed to check that the generalised Lie derivative (\ref{transformationyo}) satisfies
\be
\hat{\delta}_{\xi_1}({\hat{\cal L}}_{\xi_2}V)={\hat{\cal L}_{\xi_1}}({\hat{\cal L}}_{\xi_2}V)\label{MC},
\ee
as in the $E_7\times\mathbb{R}^{+}$, this relation establishes the covariance of the generalised Lie derivative with respect to itself, we will continue calling it ``the Leibniz property''. We write the generalised transformation as
\be
(\hat{\cal L}_{\xi}V)^M=({\cal L}_{\xi}V)^M-f^{MP}{}_Q\Sigma_PV^Q
\ee
where $({\cal L}_{\xi}V)^M$ is (\ref{transformation8b}) and the failure of this to satisfy the Leibniz property is given by (\ref{failure}). Applying the $\hat{\Delta}_{\xi_1}={\hat{\cal L}_{\xi_1}}-\hat{\delta}_{\xi_1}$ operator, where $\hat{\delta}_{\xi_1}$ is defined through the relation $\hat{\delta}_{\xi_1}V=\hat{\cal L}_{\xi_1}V$, on the generalised Lie derivative ${\hat{\cal L}}_{\xi_2}V$ we get
\bea\label{closure8}
\big(\hat{\Delta}_{\xi_1}({\hat{\cal L}}_{\xi_2}V)\big)^M & = & f^{MP}{}_Q\Big(\hat{\delta}_{\xi_{1}}\Sigma_{2P}-L_{\xi_{1}}\Sigma_{2P}-\partial_K\xi_{1}^K\Sigma_{2P}\\ \nonumber
{} & {} & +f_K{}^J{}_I\partial_P\partial_J\xi_{1}^I\xi_{2}^K-\partial_P\Sigma_{1K}\xi_{2}^K\Big)V^Q.
\eea
From the above expression we can see that the Leibniz property implies \footnote{I thank Martin Cederwall for pointing me out that the Leibniz property, in the sense (\ref{MC}), can be achieved rather than only closure of the algebra, and therefore covariance. Also for sharing his unpublished notes with me and specially for let me use his result (\ref{closure8}) concerning the Leibniz property in the $E_8$ case.} (notice no antisymmetrization is needed)
\be\label{sigmavariation}
\hat{\delta}_{\xi_{1}}\Sigma_{2P}=L_{\xi_{1}}\Sigma_{2P}+\partial_K\xi_{1}^K\Sigma_{2P}-f_K{}^J{}_I\partial_P\partial_J\xi_{1}^I\xi_{2}^K
+\partial_P\Sigma_{1K}\xi_{2}^K.
\ee
As we will see, from the compatibility with the transformation
$\hat{\delta}_{\xi_{1}}\Sigma_{2P}$ computed using that $\Sigma$ is given by (\ref{effecpara}),  (\ref{sigmavariation}) is the right expression for the $\Sigma$ transformation.

In the computation of the Leibniz property we have used the identity
\be
f^{P}{}_{M}{}^{K}f_{PN}{}^{L}C_{K}\otimes C^{\prime}_{L} = C_M\otimes C^{\prime}_N+C_N\otimes C^{\prime}_M\label{identsc}
\ee
where, in our case,
\be
C_M=(C_{\hb}{}^{\ha},0,0)=(C_{\beta}{}^{9},0,0),
\ee
 but in general the only that is required for (\ref{identsc}) to hold is that $C$ and $C^{\prime}$ are solutions of (\ref{seccon8}), which implies that the Leibniz property holds for any solution of the section condition. This identity has been proven in \cite{Hohm:2014fxa}, also one can see that the identity is a consequence of the last line of (\ref{seccon8}) which can be proven from (\ref{estructure}).
Another useful identity that follows from (\ref{identsc}) is
\be
{\cal L}_{f^{\bullet}{}^N{}_P\Sigma_N{}^P}V^M=f^{IM}{}_J(\partial_I\Sigma_P{}^P+\partial_P\Sigma_I{}^P)V^J
\ee
where, in our case,
\be
\Sigma_N{}^P=(\Sigma_{\hb}{}^{\ha}{}^{\ P},0,0)=(\Sigma_{\beta}{}^{9}{}^{\ P},0,0).
\ee
\subsection*{Compatibility}

If $\Sigma$ were an independent parameter we should check that the Leibniz property also holds to this new component of the generalised vector, i.e we should check that
\be
\hat{\delta}_{\xi_1}({\hat{\cal L}}_{\xi_2}\Sigma)={\hat{\cal L}_{\xi_1}}({\hat{\cal L}}_{\xi_2}\Sigma)\label{MC1},
\ee
as usually this checking is performed in the tensor hierarchy mechanism \cite{Aldazabal:2013via}. However as $\Sigma$ actually is not an independent parameter the consistency check becomes a compatibility check.

Having a notion of generalised Lie derivative it is possible to define the generalised covariant derivative $\nabla$. In general for some generalised connection $\Gamma$ it is written in an local generalised patch as
\be
\nabla_MV^N=\partial_MV^N+\Gamma_{MK}{}^NV^K.
\ee
Demanding that $\nabla_MV^N$ transforms as a tensor with respect to the generalised transformation (\ref{transformationyo}) we get
\be
\hat{\delta}_{\xi_1}\Gamma_{MK}{}^N=\hat{\cal L}_{\xi_1}\Gamma_{MK}{}^N+f_A{}^N{}_Kf^{AP}{}_Q\partial_M\partial_P\xi_{1}^Q+f^{NP}{}_K\partial_M\Sigma_{1P}
-\partial_M\partial_P\xi_{1}^P\delta_K^N\label{gammatransf}
\ee
where $\hat{\cal L}_{\xi_1}\Gamma$ is denoting the tensorial part of the $\Gamma$ transformation.
Using
\be
\nabla_ME_{\bar A}{}^N=W_{M\bar{A}}{}^{\bar{B}}E_{\bar{B}}{}^N=\partial_ME_{\bar A}{}^N+\Gamma_{MK}{}^NE_{\bar A}{}^K.
\ee
where $W_{M\bar{A}}{}^{\bar{B}}$ is the generalised spin connection (notice that it is defined as in the Riemann geometry but $\nabla_M$ is the generalised covariant derivative) it is easy to prove that
\be
(\hat{\delta}_{\xi_1}-\hat{\cal{L}}_{\xi_1})\Gamma_{MK}{}^N=-(\hat{\delta}_{\xi_1}-\hat{\cal{L}}_{\xi_1})\Omega_{MK}{}^N\label{deltagamma}.
\ee
Using (\ref{deltagamma}) and (\ref{gammatransf}) we get
\be
\hat{\delta}_{\xi_1}\Omega_{MK}{}^N=\hat{\cal L}_{\xi_1}\Omega_{MK}{}^N-f_A{}^N{}_Kf^{AP}{}_Q\partial_M\partial_P\xi_{1}^Q-f^{NP}{}_K\partial_M\Sigma_{1P}
+\partial_M\partial_P\xi_{1}^P\delta_K^N\label{omegatransf}.
\ee
In particular (\ref{omegatransf}) also holds for $\tilde{\Omega}_{MK}{}^N$. Now, contracting with $f_I{}^K{}_N\xi_{2}^I$, using (\ref{cartan}) and $f_I{}^K{}_K=0$
we get
\be
\hat{\delta}_{\xi_1}(\frac{1}{60}f_I{}^K{}_N\tilde{\Omega}_{MK}{}^N\xi_2^I)=\hat{\cal L}_{\xi_1}(\frac{1}{60}f_I{}^K{}_N\tilde{\Omega}_{MK}{}^N\xi_2^I)-f_I{}^{P}{}_Q\partial_M\partial_P\xi_{1}^Q\xi_2^I+\partial_M\Sigma_{1I}\xi_{2}^I\label{prove},
\ee
or, from (\ref{effecpara})
\be
\hat{\delta}_{\xi_1}\Sigma_{2M}=\hat{\cal L}_{\xi_1}\Sigma_{2M}-f_I{}^{P}{}_Q\partial_M\partial_P\xi_{1}^Q\xi_2^I+\partial_M\Sigma_{1I}\xi_{2}^I\label{prove1},
\ee
where, due to the fact that $\Sigma$  has zero weight, we have
\be
\hat{\cal L}_{\xi_1}\Sigma_{2M}={\cal L}_{\xi_1}\Sigma_{2M}=L_{\xi_1}\Sigma_{2M}+\partial_P\xi_1^P\Sigma_{2M}.
\ee
Notice that (\ref{prove1}) is exactly the same expression (\ref{sigmavariation}) we computed through the Leibniz property, therefore our check of consistency and compatibility has been achieved. We want to stress that in the computation of the Leibniz property as well as in the compatibility check we have only used the fact that all fields satisfy the section condition (\ref{seccon8}) and the computation has been performed without assuming the particular embedding (\ref{emb1}) and (\ref{emb2}). Therefore, our analysis is also valid for other solutions of the section conditions in particular the one relevant in type IIB supergravity presented in \cite{Hohm:2014fxa}.

\section{Eleven dimensions}\label{e11}

\subsection{From $E_8\times\mathbb{R}^{+}$ to eleven dimensions}

To go further from eight dimensions we could take as the starting point the generalised Lie derivative presented above. Before we discuss this statement we will expose a few facts about the generalised Lie derivative on the groups $E_d\times\mathbb{R}^{+}$, with $d\leq7$.

On the one hand, the three first lines of the generalised Lie derivative (\ref{transformation_no_indices}) satisfy the Leibniz property not only for $d\leq6$ but also for all $d$. When the algebra is extended with the fourth line of (\ref{transformation_no_indices}), the consistency of the generalised Lie derivative gets reduced to $d=7$. On the other hand, the covariance of the $E_8\times\mathbb{R}^{+}$ generalised Lie derivative does not depend on the dimension of the manifold, hence this transformation satisfies the Leibniz property for all $d$. However, we have to be very careful, since the transformation depends on having a well defined generalised parallelisable manifold and for $d>8$ we can not claim its existence since these kind of manifolds have not been studied yet.

For $d>8$ the U-duality groups are infinite dimensional, thus a generalised frame $E_{\bar{A}}{}^P$ on these groups would be an infinite set of infinite-dimensional vectors. To write explicitly the full Weitzenb\"{o}ck connection $\Omega_{MN}{}^P=E_N{}^{\bar{A}}\partial_ME_{\bar{A}}{}^P$ in terms of the degrees of freedom  would be trickier, since an infinite sum is involved in its definition. Regarding this fact, recently there appeared two interesting papers
\cite{Tumanov:2014pfa}, \cite{West:2014eza} which based on them could be worth exploring how to go beyond eight dimension  within the approach displayed in this work.

Finally, we are confident that under the $SL(d)$ decomposition and proper truncation (in particular the $SL(11)$ decomposition and proper truncation of $E_{11}$) an extension of  the $E_8\times\mathbb{R}^{+}$ generalised transformation to $d>8$ could be possible.

\section{Summary and Conclusions}\label{conclu}

In this work we have constructed the $E_8\times\mathbb{R}^{+}$ generalised transformation which is conceptually different to the one presented in \cite{Hohm:2014fxa}. Remarkably,  its consistency is not subject to any compensating parameter, thus only the parameter and the degrees of freedom of the $E_8\times\mathbb{R}^{+}$ group are involved in the transformation. Although no compensating fields are needed, the generalised Lie derivative seems not to have a covariant coordinate-independent writing. This could be a problem for the covariance of the theory. However, when the theory is defined on a generalised parallelisable manifold a consistent transformation is achieved upon the introduction of the generalised Weitzenb\"{o}ck connection.

The extended $E_7\times\mathbb{R}^{+}$ generalised approach was used as a laboratory. In particular, we present the $SL(8)$ and $SL(7)$ decomposition of the extended generalised $E_7\times\mathbb{R}^{+}$ transformation, obtaining perfect agreement with \cite{Coimbra:2012af}. From the $SL(7)$ perspective we computed the consistency conditions, which indeed are the closure of the algebra and the Leibniz identity, and we  analysed under what conditions the transformation is consistent. As expected, in seven dimensions there is no problem with the generalised Lie derivative. However, beyond seven dimensions the Leibniz property does not hold, hence an extension from $d=7$ is impossible.

Working out explicitly the  $E_7\times\mathbb{R}^{+}$ fluxes definition, it was possible to show that they can be written as a combination of  the projectors to the ${\bf 56 + 912}$ irreducible representations of the $E_7$ group acting on the Weitzenb\"{o}ck connection, plus terms associated with the conformal factor. In fact, it is valid for all $E_d\times\mathbb{R}^{+}$ with $ d\leq7$, where now the projection is on the correspondent ${\bf R_1+R_3}$ irreducible representations of the $E_d$ group. Interestingly enough, the same expression can be written for $d=8$, but the only difference with the other exceptional groups is that ${\bf R_3}$ is a reducible representation of $E_8$.

Using the lessons learnt from the $E_7\times\mathbb{R}^{+}$ case, we moved forward to the $E_8\times\mathbb{R}^{+}$ one. We presented the full $d=8$  generalised transformation, written in terms of  the fundamental indices of the  $E_8$ group  and splitted in indices of the $SL(8)$ one. Consistency and compatibility were checked, showing that the transformation of $\Sigma$ can be computed through the consistency conditions or using its own definition in terms of the vector parameter and the Weitzenb\"{o}ck connection. Actually, this is a non trivial statement and its checking strengthens the arguments presented here.

We shall now stress some facts about what is called Exceptional Field Theory (or its counterpart Double Field Theory) and Generalized Geometry. EFT and DFT
are formulated in a coordinate-dependent way. This does not mean that they do not make sense. It is well known that upon solving, properly,  the section condition, locally these kind of theories are equivalent to generalised geometry, which is a well defined and covariant theory, whose consistency does not depend on any choice of coordinates.

In this paper we restrict our attention to give, as good as possible, a definition of generalised Lie derivative for the $E_8$ generalised geometry, restricted to eight dimensions. Then, based on the fact that the consistency of the generalised Lie derivative presented here does not depend on the dimension of the manifold,  we conjectured that this derivative could be taken as the starting point to describe the full geometry of the M-theory in eleven dimensions.

The attempt of \cite{Hohm:2014fxa} was to describe the geometry of the eleven dimensional M-theory, however, the starting point, there, was a generalised Lie derivative which is not covariant, which means that the approach only works locally. Here we presented a refined version of the generalised Lie derivative presented in \cite{Hohm:2014fxa}. To avoid some consistency issues that appears in the $E_8$ case a new local parameter was introduced in \cite{Hohm:2014fxa}, the point is that this new parameter gives rise a new and undesirable local degree of freedom.  As in our approach  no new parameter is needed, no further degree of freedom is needed to have a well defined gauge transformation, thus the above conjectured extension to eleven dimensions will not need any further parameter neither any further degree of freedom to be well defined.

We want to emphasize that the generalised Lie derivative presented here seems not to be a covariant object, unless, the manifold is a generalised parallelisable one. However if the attention is restricted to this kind of backgrounds then  it is possible to give a good definition of generalised Lie derivative with only the field content that corresponds to the one that the $E_8$ group admits.

There are several unanswered questions. Maybe, the two ones who need to be  imperatively answered  are,

\begin{itemize}
\item Is it possible to go beyond the generalised parallelisable manifold?
\item Is it possible to get a coordinate-independent writing of the fourth component of the $E_8\times\mathbb{R}^{+}$ generalised transformation?
\end{itemize}

Probably the answer to the first question gives us a clue to answer the second one. The only place where we used the information, apart from the flux definition, that the manifold has to be a generalised parallelisable one is in the compatibility check. Notice that (\ref{prove1}) only needs (\ref{omegatransf}) which indeed is (\ref{gammatransf}). Thus given a general connection on a general manifold seems to be sufficient for the consistency conditions to hold. The point is that having a general connection with the proper transformation is just a necessary condition. One requirement for the consistency of the generalised Lie derivative is that $\Sigma_P$ has to satisfy (\ref{embeding}). In terms of a generalised connection, $\Sigma$ should be
\be
\Sigma_{P}=\frac{1}{60}f_J{}^K{}_L\Gamma_{PK}{}^L\xi^J\label{gammasigma}.
\ee
The above expression implies that to give an $E_8\times\mathbb{R}^{+}$ generalised transformation on a general manifold, this one has to be equipped with a non zero and section-projected in its first index generalised connection. Concretely,
\be
\nabla_M=(\nabla_{\ha}{}^{\hb},0,0)=(\nabla_{\alpha}{}^{9},0,0)=(\nabla_{\alpha},0,0)
\ee
or
\be
Y^M{}_N{}^P{}_Q\nabla_M\otimes\nabla_P=0\label{SSC}.
\ee
This means that the section condition should be extended to the covariant derivatives. The relation (\ref{SSC}) would have strong implications on the definition of the generalised torsion, curvature and Ricci tensors.

The answer to the second question is more elusive. One possibility could be that on a  general manifold the generalised transformations takes the following form
\be
(\hat{\delta}_{\xi}V)^M=\xi^P\nabla_PV^M-f_A{}^{M}{}_Nf^{AP}{}_Q\nabla_P\xi^QV^N+\nabla_P\xi^PV^M\label{transfornabla},
\ee
for some general generalised connection. Then by some mechanism or imposing certain conditions like the torsion free one, (\ref{SSC}) etc; (\ref{transfornabla}) reduces to (\ref{transformationyo}). However, the meaning of these constraints as well as the definition of the torsion tensor, in this context, in not clear for us yet.

\section*{Acknowledgments}
We thank G. Aldazabal, Martin Cederwall,  M. Gra\~na, S. Iguri, D. Marqu\'es and D. Waldram for useful discussion comments and support. Particularly,  I would like to thank D. Marqu\'es for his collaboration in early stages of this work. We also thank  D. Waldram for his comments and many suggestions about the generalised transformation and connections. We specially thank Martin Cederwall for sharing his unpublished notes concerning the covariance of the generalised Lie derivative. This work was partially supported by EPLANET, grant PICT-2012-513, the ERC Starting Grant 259133 ObservableString and postdoctoral scholarship of Consejo Nacional de Investigaciones Cient\'ificas y T{\'e}cnicas (CONICET), Argentina. I would like also to thank the Particles and Fields Group of the Centro At\'omico de Bariloche for hospitality and support during the completion of this work.

\end{document}